\documentclass[aps,prd,preprint,superscriptaddress,showpacs]{revtex4}
\usepackage{graphics}
\usepackage{epsfig}
\usepackage{latexsym}
\usepackage{colordvi}
\usepackage{amsmath}
\usepackage{amssymb}

\begin{document}
\preprint{\parbox[b]{1in}{ \hbox{\tt PNUTP-05/A02} \hbox{\tt MITCTP-3654} }}

\title{RG analysis and Magnetic instability in gapless superconductors}

\author{Deog Ki Hong}
\email[E-mail: ]{dkhong@pusan.ac.kr, dkhong@lns.mit.edu}
\affiliation{Department of Physics, Pusan National University,
             Busan 609-735, Korea}

\affiliation{Center for Theoretical Physics, Massachusetts
             Institute of Technology, \\ Cambridge, MA 02139, USA}

\vspace{0.1in}

\date{\today}

\begin{abstract}
We study the magnetic instability of gapless superconductors.
The instability
arises due to the infrared divergence associated with the gapless modes
of the superconductor. When the Fermi-surface mismatch between pairing fermions
is close to the gap, the gapless modes have a quadratic energy dispersion
relation at low energy and open a secondary gap at the Fermi surface,
which is  only power-suppressed by the  coupling.
On the other hand,  for a large mismatch, we find the gapless
superconductor does not open a secondary gap, but instead makes
transition to a new phase by forming the condensate of supercurrents.
We calculate the condensate
of supercurrents by minimizing the effective potential.
In the new phase,
the Meissner mass is positive for the magnetic fields orthogonal to the direction of
the condensate but zero in the parallel direction.
\end{abstract}
\maketitle

\newpage

How matter behaves at extreme density is ultimately related to the
fundamental questions on basic building blocks of matter.
Quantum chromodynamics (QCD), which is believed to be the
theory of strong interaction of elementary particles,
predicts matter at extreme density is
color superconducting quark matter,
where the quarks form Cooper-pairs~\cite{Barrois:1977xd}.
The search for quark matter
in heavy ion collision or in compact stars
is currently under intense investigation.

Quark matter is expected to exhibit a rich phase structure, having
various pairing patterns for quarks, as temperature or density changes.
At high density, where the $\rm SU(3)$ flavor symmetry among $u,d,s$ quarks
is good, the QCD interaction pairs quarks to form color-flavor-locked (CFL)
matter, respecting the flavor symmetry~\cite{Alford:1998mk}.
The flavor-asymmetric electroweak interaction and quark mass
strain the CFL matter, which then becomes unstable under a large stress.

BCS pairing 
breaks in general when the stress is bigger than
the pairing energy or $\delta\mu>\Delta_0/\sqrt{2}$,
where the stress $2\,\delta\mu$ is the chemical potential difference of pairing quarks
and $\Delta_0$ is the BCS gap~\cite{clogston}.
Recently, however, it has been shown that Cooper-pairing may be stable even if the
stress is bigger than the gap, known as Sarma phase~\cite{Sarma},
if one enforces the electric neutrality in quark
matter~\cite{Shovkovy:2003uu} or uses a momentum-dependent kernel
for pairing~\cite{Forbes:2004cr}.
The salient feature of such asymmetric 
quark matter is that quarks can be excited at arbitrarily low energy,
though it is superconducting.
The gapless superconductors, if realized, will be a new kind of Fermi liquid,
which has both properties of the usual BCS superconductor
and the Landau-Fermi liquid.
It was soon found, however, that the gapless superconductors have negative
Meissner mass squared~\cite{Huang:2004bg} or a negative supercurrent
density~\cite{Wu-Yip}, showing magnetic instability of the system.

In this letter we calculate the effective potential for the gapless superconductors
and study the magnetic instability.
We find that the magnetic instability arises
because of the infrared divergences associated with the gapless modes.
When the stress is greater than the gap, $\delta\mu\gnsim\Delta$,
the infrared divergence is logarithmic and the system develops a
condensate of supercurrents.
On the other hand,
when the stress is very close
to the gap, $\delta\mu\approx\Delta$, the divergence is much more severe
and it inevitably leads to opening a secondary gap
at the Fermi surface. 

To study the secondary gap in detail,
we derive the low energy effective Lagrangian for the
gapless modes by integrating out the gapped modes, which turns out to have
an attractive four-Fermi interaction.
Unlike the ordinary Fermi liquid,
the four-Fermi interaction in gapless superconductors
scales like $s^{-1/2}$ as one scales down to the Fermi surface ($s\to0^+$),
if incoming fermions have equal and opposite momenta~\cite{Liu:2004mh}, and
induces a gap which is only power-suppressed in couplings.
Both renormalization group (RG) analysis and gap equation analysis
show that the gapless excitations develop a gap, which naturally
stabilizes the system.

Let us consider a minimal model for gapless superconductivity,
which has two different flavors that pair, denoted as up and down, and
free electrons for electric neutrality, shown in Table~\ref{qn}.
The pairing force is $SU(2)_c$ color interaction under which
the up and down particles are fundamental.
\begin{table}[htb]
\begin{center}
\begin{tabular}{|c||c|c|c|}
\hline  & chemical potential & electric charge & $SU(2)_c$\\ \hline
$\psi_1$ (up) & $\mu_1=\bar\mu-\delta\mu$ & $q_1=\bar q+\delta q$ & 2\\ \hline
$\psi_2$ (down) & $\mu_2=\bar\mu+\delta\mu$ & $q_2=\bar q-\delta q$ & 2\\ \hline
$\psi_e$ (electron) & $\mu_e=2\delta\mu$ & $q_e=-1$ & 1\\ \hline
\end{tabular}\end{center}
\caption{The electron chemical potential is $\mu_e=\mu_2-\mu_1$ and
$q_e=-2\delta q$, since
the system is in equilibrium under the weak interaction,
$\psi_2\leftrightarrow\psi_1+e^-+\bar\nu_e$.
}\label{qn}
\end{table}
The system is not a color superconductor, since the Cooper pair is a color-singlet.
However, it is an electric superconductor and exhibits all the essential features
of gapless superconductivity, including the magnetic instability. It is
straightforward to extend to the gapless 2-flavor color-superconductor (g2SC)
or to the gapless CFL superconductor (gCFL).
Neglecting the anti-particles, the system is described by a Lagrangian density
\begin{eqnarray}
{\cal L}=\sum_{i=1,2}\psi_i^{\dagger}\left[i\partial_t-E(\vec p\,)+\mu_i\right]\psi_i
+\frac{G}{2}\,\epsilon^{ij}\psi_{i}^{\dagger}\psi_{j}^{\dagger}
\epsilon^{i^{\prime}j^{\prime}}
\psi_{i^{\prime}}\psi_{j^{\prime}}
+{\cal L}_{\rm int}^{\prime}\,,
\end{eqnarray}
where
$E(\vec p\,)$ is a flavor-independent function of momentum and
the color and spin indices are suppressed. Gluons are integrated out
to generate an attractive four-Fermi interaction,
antisymmetric in color and flavor, and also other irrelevant interactions, denoted as
${\cal L}_{\rm int}^{\prime}$, assuming
the gluon exchange interaction is most attractive for quarks
antisymmetric in color and flavor.

Neglecting the irrelevant interactions,
one can calculate the free energy
in the mean field approximation,
\begin{eqnarray}
\Omega_s(\Delta,\delta\mu)=\frac{\Delta^2}{G}-2
\int\!\frac{{\rm d}^3p}{(2\pi)^3}\left[\sqrt{\epsilon^2+\Delta^2}+\delta\mu
+\left|\sqrt{\epsilon^2+\Delta^2}-\delta\mu\right|\,
\right]-\frac{(2\delta\mu)^4}{12\pi^2}\,,
\label{free_energy}
\end{eqnarray}
where $\epsilon(p)=E(p)-\bar\mu$ and the last term is the electron free energy.
When $\delta\mu>\Delta$, the gap equation has an unstable
solution, $\Delta=\sqrt{\Delta_0\left(2\delta\mu-\Delta_0\right)}$,
where $\Delta_0$ is the gap when $\delta\mu=0$.
This unstable Sarma phase can be stabilized
if charge neutrality is enforced.
The electric charge neutrality condition for the Sarma phase,
$\left<Q_{\rm em}\right>=0$,
is satisfied for $\bar\mu\gg\delta\mu$ if
$\delta\mu=\sqrt{\delta\mu_0^2+\Delta^2}\,$,
where $2\,\delta\mu_0$ is the chemical potential difference of
the free neutral system~\cite{Huang:2003xd}.
Compared with the free neutral system,
the free energy becomes for $\bar\mu\gg\delta\mu$
\begin{eqnarray}
\Omega_s(\Delta,\delta\mu)-\Omega_{\rm free}
\simeq-\frac{{\bar\mu}^2}{\pi^2}\,\delta\mu_0^2
\left[1-\left(\frac{\Delta_0}{2\delta\mu_0}-1\right)^2\right]\,.
\end{eqnarray}
The neutrality enforcement stabilizes the Sarma phase if
$\Delta_0/4\lesssim\delta\mu_0\lesssim\Delta_0/2$
or the gap $0\lesssim\Delta\lesssim\Delta_0/\sqrt{2}$.
At the minimal $\delta\mu_0$,
the ratio $\delta\mu/\Delta$ is smallest and close to $7/4-1/\sqrt{2}$.

We now calculate the Coleman-Weinberg effective
potential~\cite{Coleman:1973jx} for photon fields
in the Sarma phase, which will be
same as the free energy in Eq.~(\ref{free_energy}) except that
$p_0\to p_0+q_i\,eA_0$ and $\vec p\to \vec p+q_i\,e\vec A$~\cite{Weinberg:1986cq}.
As we integrate out the high frequency modes, the effective potential will
get non-local terms, which can be expanded in powers of momentum.
When all the modes with $\omega\gtrsim\delta\mu$ are integrated out,
only the gapless modes survive and the effective potential becomes
\begin{eqnarray}
V(A)\!=\!
-4\!\int_{\Lambda}\!\frac{{\rm d}^4p_E}{(2\pi)^4}
\ln\!\left[p_4^2+\left(
\sqrt{\bar\epsilon_1\bar\epsilon_2+\Delta^2}
-\delta\bar\mu\right)^2\right]+C_0(\Lambda)+\frac{C_2(\Lambda)}{2}\!
{\vec A\,}^2\!-\!\frac{B_2(\Lambda)}{2}{A_0}^2\!+\!\cdots
\end{eqnarray}
where
$\delta\bar\mu=\delta\mu-\delta q\,eA_0$ and
$\bar\epsilon_i=E(\vec p+q_i\,e\vec A\,)-\bar\mu-\bar q\,eA_0$.
In the effective potential
we have introduced an ultra-violet cutoff $\Lambda$ ($\lesssim\delta\mu$)
and the counter-terms, $C_i$'s
and $B_i$'s. The ellipsis denotes the higher order terms in photon fields
and their derivatives.

As we will see, the gapless superconductor suffers instability due to the
infrared divergences associated with the gapless quasiparticle modes.
Depending on
how close the stress is to the gap, the structure of divergences
and their physical consequences differ drastically.
When the stress is not too close to the gap ($\delta\mu\gnsim\Delta$),
as in g2SC,
the quasi-particles have approximately linear dispersion relation
near the Fermi surface. In this case
there is a logarithmic infrared divergence and the system is
stabilized by spontaneously generating Goldstone currents.
However, when the stress is close to the gap ($\delta\mu\approx\Delta$),
as in gCFL, the quasiparticles near the Fermi surface
have approximately quadratic dispersion relation
and there is genuine instability in the system,
which leads to opening a secondary gap at the Fermi surface.

We consider first the case where the stress is not too close to the gap.
If we integrate out further  the quasi-particle modes
till the remaining has an
approximately linear dispersion relation,
the effective potential becomes
\begin{eqnarray}
V(A)&=&-\nu_1\,\eta^{-1}\int\frac{{\rm d}\Omega_{\vec v_1}}{4\pi}
\left(eV_1\cdot A\right)^2\left[\ln\frac{\Lambda^2}
{\left(eV_1\cdot A\right)^2}+1\right]+(1\to2)\\ &&
+C_0(\Lambda)+\frac{C_2(\Lambda)}{2}
{\vec A\,}^2-\frac{B_2(\Lambda)}{2}{A_0}^2+\cdots\nonumber
\end{eqnarray}
where $\nu_1$ and $\nu_2$ are the density of states at the Fermi surfaces
at $p_1$ and $p_2$, respectively. At the inner and outer Fermi surfaces,
$\epsilon(p_1)=-\sqrt{\delta\mu^2-\Delta^2}$ and
$\epsilon(p_2)=\sqrt{\delta\mu^2-\Delta^2}$.
The angular integrations are over the inner and outer Fermi velocities,
$\vec v_1=\partial E/\partial {\vec p_1}$ and
$\vec v_2=\partial E/\partial \vec p_2$.
We have also defined $\eta=\delta\mu/\sqrt{\delta\mu^2-\Delta^2}$,
$V_1^{\mu}=(\eta\bar q+\delta q,\eta\bar q\,\vec v_1)$,
and $V_2^{\mu}=(\eta\bar q-\delta q,\eta\bar q\,\vec v_2)$.

We see that the one-loop effective potential due to the gapless
modes is negative for both $A_0$ and $\vec A$, and its second derivative is
negative infinity at the origin. Since the Meissner mass
term in the potential has the positive sign, while the Debye mass has a wrong
sign, the effective potential has a minimum away from the origin for
$\vec A$. 
We fix the counter-terms by imposing the renormalization conditions at a scale $M$,
\begin{eqnarray}
\left.\frac{\partial^2V}{\partial A_0^2}\right|_{{A_0}=M}\!\!=-m_D^2\,,\quad
\frac{1}{3}\,\delta_{ij}\left.
\frac{\partial^2V}{\partial A_i\partial A_j}\right|_{{\vec A\,}^2=M^2}
=m_M^2\,,
\label{r_cond}
\end{eqnarray}
to get the effective potential
\begin{eqnarray}
V(A)&=&\frac{1}{2}m_M^2{\vec A\,}^2
-\frac{\eta}{3}\left(\nu_1v_1^2+\nu_2v_2^2\right)e^2{\bar q\,}^2\,{\vec A\,}^2
\left[\ln\left(\frac{M^2}{{\vec A\,}^2}\right)+3\right]\nonumber\\
&&-\frac{1}{2}m_D^2A_0^2-\frac{1}{\eta}\left(\nu_1+\nu_2\right)\,
\left(\eta\bar q+\delta q\right)^2e^2{A_0}^2
\left[\ln\left(\frac{M^2}{{A_0}^2}\right)+3\right]\,
\label{e_pot}
\end{eqnarray}
The physics lies in the renormalization conditions that we imposed in
Eq.~(\ref{r_cond}). If we take $M=0$, we will recover the negative mass squared
for the Meissner mass, similar to the result
obtained in~\cite{Alford:2005qw}. However, this result is
sensitive to the ultraviolet cutoff. Here instead we take a nonzero $M$ such that
the Meissner mass due to modes with $\omega>M$ is nonnegative and then we study
how the system flows as we change $M$, keeping the ultraviolet physics in the
renormalization conditions.

The vector fields in Eq.~(\ref{e_pot}) are not physical, because the effective
potential we derived is not manifestly gauge-invariant. For the physical
degrees of freedom, we go to a unitary gauge, where the vector fields are replaced
by the gauge-covariant combination of gauge fields and Goldstone fields, 
$\vec A\to \vec A-\vec\nabla \varphi$, $A_0\to A_0+\partial_t\varphi$.
By minimizing the effective potential, Eq.~(\ref{e_pot}),
we find that the system develops a condensate, breaking the rotational invariance,
\begin{equation}
\left<\!\left({\vec A-\vec \nabla\varphi}\right)^2\!\right>\!
=\!M^2\exp\!\left[2\!-\!\frac{3\,m_M^2}{2\eta\left(\nu_1v_1^2+\nu_2v_2^2\right)e^2{\bar q}^2}
\right]. 
\end{equation}

Since the condensate is invariant under the renormalization group flow,
we calculate it at $M=\delta\mu$.
The Meissner mass due to the fast modes ($\omega\ge\delta\mu$) is
found to be $m_M^2\simeq(8/3)\,{\bar q}^2\,e^2\,\nu_*\,v_*^2$, where
$\nu_*$ and $v_*$ are the density of states and the quasiparticle velocity at
the pairing momentum, $p_*$, respectively. We find
\begin{eqnarray}
\left<\left({\vec A-\vec \nabla\varphi\,}\right)^2\right>
\simeq{\delta\mu}^2\,
\exp\left[2-\frac{4\,\nu_*\,v_*^2}
{\eta\left(\nu_1v_1^2+\nu_2v_2^2\right)}
\right]\,,
\end{eqnarray}
which shows the system has a spontaneously generated Goldstone current even
in the absence of external gauge fields,
$\left<-\vec \nabla\varphi\right>\equiv\vec A_c\ne0$~\cite{footnote1}.
The ground state, however, does not carry any net current, since
$\left<\vec J\,\right>\equiv
-\left.\partial V(A)/\partial \vec A\right|_{\vec A=\vec A_c}=0$~\cite{london}.

The gapless superconductor is a directionally perfect diamagnet 
because the effective potential has flat directions
along the angular rotation of  $\vec A_c$.
Under an external field $\vec A$, the current is given as
\begin{eqnarray}
J_i(x)=-\left.\frac{\partial^2V(A)}{\partial A_i\partial A_j}\right|_{\vec A=\vec A_{c}}
\left(A_j-\partial_j\varphi\right).
\end{eqnarray}
Taking the curl of the current, we get
$\nabla^2\vec B=c_M^2\left(\vec B-\vec B\cdot\hat A_c\,\hat A_c\right)$,
where where $\hat A_c$ is the unit vector along the condensate
and $c_M^2=\frac{2\eta}{3}\left(\nu_1v_1^2+\nu_2v_2^2\right)e^2{\bar q\,}^2$.
The Meissner mass is $c_M$ for the magnetic fields orthogonal
to the condensate and zero for the parallel fields.

When $\delta\mu
\approx\Delta$, $p_1\approx p_2\approx p_*$ and
the gapless modes have a quadratic dispersion relation
$\omega(\vec p\,)\simeq(\vec v_*\cdot\vec l\,)^2/(2\delta\mu)$
for $0\lesssim\omega\lesssim\delta\mu$,
where the residual momentum $\vec l=\vec p-\vec p_*$.
The effective potential due to the quadratic gapless modes is found to be,
with $V_*=(1,\vec v_*)$,
\begin{eqnarray}
\delta V(A)
=\!\!-\frac{16\Gamma(5/4)\Gamma(1/4)}{3\sqrt{\pi}}\!
\int\frac{{\rm d}\Omega_{\vec v_*}}{4\pi}
\frac{\nu_*\,|e|^3}{2\delta\mu\,|\vec v_*|}\,
\left|q_1V_*\cdot A-\delta qA_0\right|^{3/2}
\left|q_2V_*\cdot A+\delta qA_0\right|^{3/2}.
\end{eqnarray}
The effective potential shows a genuine instability,
not stabilized by the condensate of Goldstone currents,
since it is
unbounded from below. 
However, as we  will see later,
by the Kohn-Luttinger theorem~\cite{kohn_luttinger},  the quadratic gapless modes
pair among themselves to open a gap at $p=p_*$ and the system becomes stable.

The Sarma phase has a gapless excitation,
denoted as $\Psi$, whose energy spectrum is given as
$\omega=\pm\left(\delta\mu-\sqrt{\epsilon(p)^2+\Delta^2}\right)$,
and a gapped excitation, denoted as $\Psi_{\rm H}$, with
energy spectrum
$\omega=\pm\left(\delta\mu+\sqrt{\epsilon(p)^2+\Delta^2}\right)$.
The energy of the gapless mode, measured from the Fermi surface
is quadratic in the
residual momentum except very near the Fermi surface.

Subtracting out the Fermi momentum as in the high density effective
theory~\cite{Hong:2000tn},
we may write the effective Lagrangian for gapless modes as,
when $\delta\mu>\omega\gtrsim\omega_{\rm IR}\equiv(\delta\mu^2-\Delta^2)/(2\delta\mu)$,
\begin{eqnarray}
{\cal L}_{\rm eff}=
\sum_{\vec v_*}\Psi^{\dagger}
\left[i\partial_t+\frac{(\vec v_*\cdot\vec\nabla)^2}{2\,\delta\mu}\right]
\Psi(\vec v_*,x)
+\frac{\kappa}{2}\Psi^{\dagger}\Psi^{\dagger}\Psi\Psi+\cdots,
\end{eqnarray}
where the summation is over all the patches that cover the Fermi surface and
the ellipsis denotes the higher order interactions.

Because of the quadratic dispersion relation, as we scale down to the Fermi surface
$\omega\to\,s\,\omega$ ($0<s<1$), the momentum parallel to the Fermi velocity $\vec v_*$
scales as
$\vec v_*\cdot \vec l\,\to\,s^{1/2}\,\vec v_*\cdot \vec l$,
while the perpendicular momentum $\vec l_{\perp}=\vec l-
\hat v_*\,\hat v_*\cdot \vec l$
does not scale. Since the action for the kinetic term has to be scale-invariant,
the gapless mode scales in the momentum space as
$\Psi(t,\vec l\,)\,\to\, s^{-1/4}\,\Psi(t,\vec l\,)\,$.
The immediate consequence of such unusual scaling is
that the four-Fermi interaction becomes a relevant
operator 
when the incoming fermions have equal and opposite Fermi momenta.
Under the scale transformation the four-Fermi coupling transforms as
$\kappa\,\to\, s^{-1/2}\,\kappa$
and it will hit an infrared singularity unless a gap opens in the infrared region.

Among the irrelevant interactions in ${\cal L}^{\prime}_{\rm int}$, which we have
neglected so far, there is a repulsive four-Fermi interaction,
${\cal L}^{\prime}_{\rm int}\ni G_s\psi_1^{\dagger}\psi_1\psi^{\dagger}_2\psi_2$,
which is symmetric in color. This interaction
describes transition between the gapless mode and the gapped mode and
induces an attractive four-Fermi interaction for the gapless modes.
In our minimal model, the attractive channel turns out to be
a spin-1 and color-singlet channel and
the condensate takes
$\left<\Psi i\sigma_2\sigma^i\lambda^2\Psi\right>\sim\Delta_s\delta^{i3}$,
similar to the polar phase~\cite{Schafer:2000tw},
where $\sigma$'s are the Pauli matrices in the spin space and $\lambda^2$ is
the antisymmetric Pauli matrix in the color space.
By integrating out the gapped modes, we have
$\kappa=(\pi/4)\,G_s^2\,\nu_*/v_*$ at one-loop  (See Fig.~1).
\begin{figure}[h]
\vskip 0.1in
\centerline{\epsfxsize=3.5in\epsffile{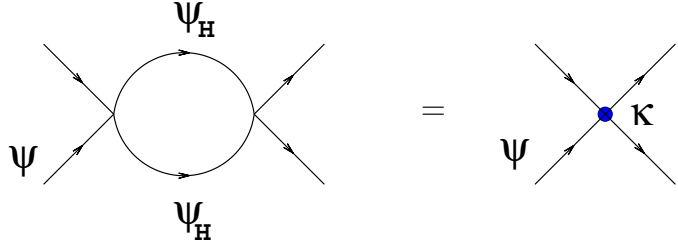}}
\caption{The four-Fermi interaction of the gapless modes $\Psi$,
induced by the gapped modes $\Psi_{\rm H}$.}
 \label{fig1}
\end{figure}

The Cooper-pair gap equation is given as
\begin{eqnarray}
\Delta_s=\kappa\,\int\frac{{\rm d}^4l}{(2\pi)^4}\frac{\Delta_s}{l_4^2+
\left[\frac{(\vec l\cdot\vec v_*)^2}{2\delta\mu}\right]^2+\Delta_s^2}.
\label{gap_equation}
\end{eqnarray}
Upon integration, we get
\begin{eqnarray}
\Delta_s\simeq6.85 \,\kappa^2\,\left(\frac{\nu_*}{v_*}\right)^2\delta\mu
=4.2\,\left(\frac{G_s}{G}\right)^2\,g^4\,\delta\mu\,,
\end{eqnarray}
where $g^{-1}\equiv\ln\,(2\bar\mu/\Delta)$\,.
For a weak coupling,
$g\ll1$,  the secondary gap is well separated from the
ultraviolet scale.
For a strong coupling, however,  one needs to go
beyond the mean field approximation to find  the correct secondary gap.
In any case, the gapless superconductors are stabilized
by opening a secondary gap at the Fermi surface if
the secondary gap is bigger than the infrared cutoff of the
effective theory, $\omega_{\rm IR}\simeq(\delta\mu^2-\Delta^2)/(2\delta\mu)$.

We have studied a minimal model for gapless superconductivity and found that
the gapless superconductors necessarily
open a secondary gap at the Fermi surface if the characteristic
scale of the gapless superconductors $\omega_{\,\rm IR}$ is small enough compared to the
secondary gap.
However,
if $\omega_{\,\rm IR}>\Delta_s$, the gapless superconductors do not open a gap but
instead develop a condensate of supercurrents.

Though our analysis is generic and applies to any gapless superconductors
or superfluids,
the specific form of the secondary gap depends on the details of the gapless
superconductors. For instance, for g2SC the gapless modes are degenerate in color and spin.
Thus, if the secondary gap forms, it will be
a color-antitriplet and spin-1 gap, but not exponentially suppressed.
For gCFL the gap has to open in the color-sextet channel,
since the gapless modes are not degenerate in color.
The possible candidate is
a color-sextet condensate,
$\left<\Psi_L\,C\gamma_0\gamma_5\Psi_R\right>$~\cite{Alford:2002rz}.


To conclude, we have shown that the gapless superconductors are unstable due to the
infrared divergence associated with the gapless modes and make phase transition
either to superconductors with a supercurrent condensate, or to gapful
superconductors. The Meissner mass is nonnegative in both cases.
In the former case, the rotational symmetry is spontaneously broken and
the Meissner mass is directional.
In the latter case the secondary gap is not exponentially suppressed but only
power-suppressed in couplings,
which may have significance in neutron stars or in atomic superfluids.


\acknowledgments
We thank M. Alford, M. Forbes, J. Goldstone, R. Jackiw, M. Mannarelli,
A. Schmitt, A. Schwenk, I. Shovkovy, and M. Stephanov for discussions.
We are thankful especially to K. Rajagopal for illuminating discussions.
This work  is supported by
Korea Research Foundation Grant (KRF-2003-041-C00073) and also in
part by funds provided by the U.S. Department of Energy (D.O.E.)
under cooperative research agreement \#DF-FC02-94ER40818.

\end{document}